\PassOptionsToPackage{numbers,sort&compress,square}{natbib}
\documentclass{iopjournal}
\usepackage{graphicx,float} 

\usepackage[english]{babel}
\usepackage{natbib}
\usepackage{amsmath, amssymb, mathtools}
\usepackage{hyperref}
\usepackage[nameinlink]{cleveref}
\Crefname{equation}{Eq.}{Eqs.}
\Crefname{figure}{Fig.}{Figs.}

\date{May 2026} 

\def\beq{\begin{equation}}
	\def\ee{\end{equation}}
\def\bi{\begin {itemize}}
\def\ei{\end{itemize}}

\def\nn{\nonumber}
\def\barv{\bar v}
\def\LL{\mathsf{L}}
\def\xib{\xi}
\def\eps{\epsilon}

\def\P{\mathsf{P}}
\def\PZ{\mathsf{P}Z}
\def\QZ{\mathsf{Q}Z}

\def\Q{\mathsf{Q}}
\def\V{\mathbf{V}}
\def\Vt{\tilde {\mathbf{V}}}
\def\Ab{\bar{\mathbf{A}}}
\def\U{\mathsf{U}}
\def\K{\mathsf{K}}

\def\mud{\mu_\textrm{d}}

\begin{document}

\articletype{Paper} 

\title{Non-negative differential mobility of one-dimensional underdamped Langevin dynamics: Proof via mapping to a marginally stable 
	oscillator}

\author{Udo Seifert
}

\affil{II. Institut für Theoretische Physik, Universität Stuttgart, 70550 Stuttgart, Germany}


\email{useifert@theo2.physik.uni-stuttgart.de}

\keywords{Langevin dynamics, differential mobility}

\begin{abstract}
For an underdamped particle driven by a constant non-conservative force across a periodic potential in one spatial dimension at finite temperature, numerical studies have shown that the differential mobility is non-negative. Here, an analytical proof will be given that is based on expressing the differential mobility by the long-time mean position of a marginally stable harmonic oscillator subject to noise in its stiffness.
\end{abstract}
\newpage
Driven Langevin dynamics in a periodic potential is arguably one of the  paradigms in non-equilibrium statistical physics. It was covered more than 40 years ago in Risken's monograph \cite{risken} and in several review articles since then \cite{reim02a,hang09,spie23}. For one-dimensional overdamped dynamics, an explicit expression for the mean velocity in terms of quadratures can be used to show that the differential mobility is positive, i.e., that with increasing non-conservative force, the mean velocity increases as well \cite{cecc96,sasa05}. Phrasing it simply, tilting a washboard potential more strongly implies a larger stationary velocity. For underdamped dynamics in one dimension, there is ample numerical evidence that the differential mobility is non-negative as well, see, e.g., \cite{risken,lato13,marc14,chen15,kepn20}. However, an  analytical proof of non-negative differential mobility (NNDM) under time-independent driving in one dimension does not seem to exist. Here, the two latter specifications are crucial since breaking them leads to violations even in the overdamped case. First, for time-dependent driving, i.e., for a force with an additional ac-component, even the absolute mobility can be negative, i.e., the mean velocity is opposite to the time-averaged force \cite{eich02,mach07}. Second, in two-dimensional potentials the differential mobility can become negative since the driving may push the particle into traps from which an escape is more difficult at a larger force \cite{cecc96,bary23}. A simple discrete model with a particle hopping along two lanes illustrates this trapping effect nicely \cite{zia02}, for generalizations, see, e.g., \cite{beni14,teza20a}. The fact that the spatial dimension enters crucially indicates that  using general operator identities may not work  for a proof of NNDM since such approaches are often insensitive to the spatial dimension. 

The purpose of this Letter is to prove that the very existence of the differential mobility implies that it cannot become negative for one-dimensional underdamped motion in a periodic potential  under a constant force. We will do so by expressing it through the long-time limit of a marginally stable harmonic oscillator that is subject to noise in its stiffness  and to initial conditions that break the spatial symmetry. To the best of my knowledge, this mapping is new. For the simpler overdamped case treated in Appendix \ref{appa}, such a mapping leads to a manifestly non-negative expression that is complementary to the established ones based on the expression for the velocity  \cite{cecc96} and on a Green-Kubo-like time integral over two-point correlation functions \cite{spec06,baie09}.

An underdamped  particle with mass $m$ in a periodic potential $V(x)=V(x+L)$ driven by a force $f$ follows the Langevin equation 
\beq m \ddot x+\gamma\dot x=f-V'(x)+\xib .\ee
The Gaussian white noise has correlations
\beq
\langle \xi(t)\xi(t'))\rangle = 2(\gamma/\beta)
\delta(t-t'),
\ee
where $\gamma$ is the friction coefficient and $\beta$ the inverse temperature.
We introduce as basic scales for time, length, and energy, $m/\gamma$, $(m/\beta\gamma^2)^{1/2},$ and $1/\beta$, respectively. Replacing all physical variables by the respective scaled versions, the
Langevin equation becomes
\beq
\dot x=v~~~\textrm{and}~~~ \dot v+ v = f-V'(x) + \eta\ee
with
\beq
\langle \eta(t)\eta(t')\rangle = 2 \delta(t-t').
\ee 
Equivalently, an initial phase point $(x_0,v_0)$ will evolve into a distribution $p(x,v,t|x_0,v_0)$ that follows  
\begin{eqnarray}
	\partial_tp(x,v,t|x_0,v_0)&=&\LL_{x,v} p(x,v,t|x_0,v_0)\\
	&=&\LL^\dagger_{x_0,v_0} p(x,v,t|x_0,v_0)
\label{eq:fp-adj}
\end{eqnarray}
with the Fokker-Planck operator
\beq
\LL_{x,v}\equiv -v\partial_x -\partial_v[-v+f-V'(x)]+\partial^2_v
\ee and its adjoint
\beq\LL^\dagger_{x,v}\equiv v\partial_x+[-v+f-V'(x)]\partial_v + \partial^2_v.
\label{eq:adj}
\ee For a smooth potential $V(x)$ and periodic boundary conditions in $x$,
we will assume (i) that there is a unique stationary distribution
$p^s(x,v;f)$  that obeys
\beq
\LL_{x,v}p^s(x,v;f)=0,
\ee  (ii) that any initial distribution will converge to this distribution
exponentially in time, and (iii) that  $p^s(x,v;f)$ depends smoothly  on $f$. A rigorous proof of these properties, which can be taken for granted from a physics perspective, is beyond the scope of this Letter.
Averages with respect to this
distribution will be denoted by $\langle ...\rangle^s$. In terms of the mean velocity, which will be written as $\bar v(f)\equiv\langle v\rangle^s$, the differential mobility is then given by
\beq
\mud(f)\equiv \partial_f\bar v(f)=\int dx\int dv~ v\partial_fp^s(x,v;f).
\ee In the following, we will notationally suppress the $f$-dependence.

Since there is no analytical expression for the stationary distribution, proving NNDM directly seems to be hard. We therefore adapt a strategy akin to the approaches of \cite{lind16,dech23a}. By multiplying (\ref{eq:fp-adj}) with $v-\barv$ and integrating over $x$ and $v$, we get
\beq
\partial_t\langle v(t)-\barv|x_0,v_0\rangle =  \LL^\dagger_{x_0,v_0} \langle v(t)-\barv|x_0,v_0\rangle,
\ee 
where $\langle ...|x_0,v_0\rangle$ denotes an average over the noise at fixed initial values.
We further integrate over $t$ from 0 to $\infty$ which yields
\beq-v_0 + \barv = \LL^\dagger_{x_0,v_0} \int_0^\infty dt \langle v(t)-\bar v|x_0,v_0\rangle. 
\label{eq:ini}
\ee
Since the integrand decays exponentially in time, the integral on the right-hand side exists. Likewise, on the left-hand side, the upper limit has not left any contribution.
We now operate on (\ref{eq:ini}) with $\partial_f$ from the left, which yields
\beq
\partial_f\barv =[\partial_{v_0} +  \LL^\dagger_{x_0,v_0}\partial_f] 
 \int_0^\infty dt \langle v(t)-\bar v|x_0,v_0\rangle.
\ee Finally, by multiplying this equation from the left with $p^s(x_0,v_0)$ and integrating over $x_0$ and $v_0$ -- which has no effect on the left-hand side --, the term involving $\LL^\dagger_{x_0,v_0}$ vanishes. Thus, we get for the differential mobility  the expression
\beq \mud= \langle \partial_{v_0} \int_0^\infty dt \langle v(t)-\bar v|x_0,v_0\rangle \rangle^s .
\label{eq:R1}
\ee While the inner average is the one over noise, the outer one runs over the initial values in the stationary state.
%

We can transform this expression by considering two trajectories with the same noise history and slightly different initial conditions. Let $x(t)$ and $x_\eps(t)\equiv x(t) + \eps y_\eps(t)$ be two such trajectories with the initial values $(x_0,v_0)$ and $(x_0,v_0+\eps)$, respectively. Their difference $\eps y_\eps(t)$ follows the deterministic dynamics
\beq
\eps \ddot y_\eps(t) + \eps \dot y_\eps(t) + V'(x(t)+\eps y_\eps(t))=0
\label{eq:ho0}
\ee with the initial conditions $y_\eps(0)=0$ and $\dot y_\eps(t)=1$.
 For a fixed trajectory $x(t)$, i.e., for fixed initial values and for a fixed noise history,  and for the moment replacing the upper limit $\infty$ of the integral in $(\ref{eq:R1})$
by a large time $T$, the derivative of this integral  becomes
\beq 
\partial_{v_0} \int_0^T dt [v(t)-\bar v]= \lim_{\eps\to 0}(1/\eps)\int_0^T \eps \dot y_\eps(t) dt= \int_0^T \dot y(t) dt ,
\ee 
where $y(t)$ follows the equation obtained from linearizing (\ref{eq:ho0}). Thus, for a fixed original trajectory $x(t)$ in the periodic potential, the trajectory $y(t)$ is the one of a deterministic damped harmonic oscillator
\beq\ddot y(t) + \dot y(t) + k(t) y(t) =0 
\label{eq:ho}
\ee 
with the initial conditions
\beq
y(0)=0~~~\textrm{and}~~~ \dot y(0)=1. 
\label{eq:ini-y}\ee
Its time-dependent   stiffness
\beq k(t)= V''(x(t))
\label{eq:k}
\ee is given by the instantaneous curvature of the original periodic potential along the trajectory $x(t)$, which may well be negative along some intervals in time. 

Performing now the average over the noise histories and over the initial values $(x_0,v_0)$, the expression (\ref{eq:R1}) for the differential mobility becomes the limit
\beq \mud=\lim_{T\to \infty} \int_0^T dt\langle \dot y(t)\rangle =
\lim_{T\to \infty} \langle y(T) \rangle 
\label{eq:R3}
\ee with the initial condition (\ref{eq:ini-y}).
From the perspective of $y(t)$, the average $\langle ... \rangle$ is over the original stationary process that here governs the fluctuating stiffness (\ref{eq:k}). This expression for the diffential mobility is our first main result. 

In general, an oscillator that follows (\ref{eq:ho}) and (\ref{eq:ini-y}) for an arbitrary bounded stationary process $k(t)$ will either have a long-time mean that vanishes -- if $k(t)$ is ``dominantly'' positive --, or the limit will not exist -- if $k(t)$ has ``too many'' negative contributions which renders the oscillator unstable. However, for a $k(t)$-process that arises from the stationary state of a driven particle in a periodic potential, the fact that all steps leading to the expression (\ref{eq:R3}) for the differential mobility have been exact implies that the $k(t)$-statistics is such that this limit of marginal stability exists. What remains to be shown is its non-negativity.

At this point, it is instructive to consider simple cases for which the averaging over the $k(t)$-process is not even necessary since all trajectories $x(t)$ lead to the same long-time limit of $y(t)$.  First, in the absence of a potential, i.e., for $k(t)\equiv 0$, $y(t)=1-e^{-t}$ solves (\ref{eq:ho}) with the initial conditions (\ref{eq:ini-y}) independently of $(x_0,v_0)$ and noise history.  Hence, (\ref{eq:R3}) implies $\mud=1$ for all $f$ as it should be. Second, for a non-periodic potential with $V''(x)>0$ throughout and $V(x)$ increasing sufficiently strongly for $x\to \pm \infty$ such that it remains binding for all $f$, $\lim_{t\to \infty}y(t)=0$ due to  damping, which implies $\mud=0$.
Indeed, in such a binding potential, the mean velocity vanishes for all  $f$. 

As a non-trivial result, we can now easily prove $\mud\geq 0$ for any periodic potential with $V''(x)<1/4$, i.e., with $k(t)<1/4$.
 With the transformation
$
y(t)=e^{-t/2} z(t)
$, the equation of motion (\ref{eq:ho}) becomes
\beq\ddot z(t) +[k(t)-1/4] z(t)= 0.
\ee The boundary conditions $z(0)=0$ and $\dot z(0)=1$ imply  $z(t)>0$ for all $t>0$ even trajectory-wise, i.e., for fixed $x(t)$. Inserting $y(t)>0$ into (\ref{eq:R3}) implies $\mud\geq 0$. In such ``weakly convex'' potentials, a trajectory that starts at the same $x_0$ and with a larger initial velocity will stay ahead of the other one for all times, ultimately  implying a non-negative differential mobility.  
%

For an arbitrary periodic potential $V(x)$,
it is necessary to average over the $k(t)$-process, which we write as
\beq k(t) = \bar k +\kappa(t)~~~\textrm{with}~~~ \bar k\equiv \langle V''(x)\rangle^s\ee in terms of its  mean $\bar k$ and its fluctuations with
$\langle \kappa(t)\rangle =0$. 
We combine position and velocity of the harmonic oscillator into a two-dimensional vector $Y(t)\equiv (y(t),\dot y(t))^{\top}$. Its equation of motion reads
\beq\dot Y = \mathbf{A} Y=[\Ab + \mathbf{V}(t)]Y \ee
with the deterministic and fluctuating contributions
\beq
\Ab \equiv \begin{pmatrix}
	0 & 1 \\
	-\bar k & -1
\end{pmatrix}~~~\textrm{and}~~~ \mathbf V(t)\equiv \kappa(t) \begin{pmatrix}
0 & 0 \\
-1 & 0
\end{pmatrix}\equiv \kappa(t)\V_0,
\label{eq:V}
\ee
respectively. With the transformation to the interaction picture, 
\beq
Z(t)\equiv e^{-\Ab t} Y(t) ,
\ee we get
\beq\dot Z(t) = \Vt(t)Z(t)
\label{eq:Z}
\ee with 
\beq\Vt(t)\equiv e^{-\Ab t} \mathbf{V}(t) e^{\Ab t}. 
\ee

The mean behavior can be extracted after introducing a projection operator $\P$ that will average any functional  of the noise history  $\Phi[\kappa(t)]$
according to $\P \Phi[\kappa(t)]=\langle \Phi[\kappa(t)] \rangle.$ Its complement is $\Q\equiv {1}-\P$. Moreover, we get
\beq
\P\Ab=\Ab \P ~~~\textrm{and}~~~ \P\Vt\P = 0 ,
\ee using, for the latter, $\langle \kappa(t)\rangle = 0$.
We apply $\P$ to (\ref{eq:Z}) and use that $\P$ commutes with $\partial_t$ to get
\beq
\partial_t \PZ=P\Vt(\PZ+\QZ)= \P\Vt\QZ.
\label{eq:PZ}
\ee
Likewise, by applying $\Q$ to (\ref{eq:Z}), we get
\beq
\partial_t \QZ=Q\Vt(\PZ+\QZ)=\Vt\PZ+ \Q\Vt\QZ.
\label{eq:QZ}
\ee

The strategy will be to solve the last equation (formally) for $\QZ$ in terms of $\PZ$ and to use the result in (\ref{eq:PZ}) to get a closed equation of motion for $\PZ$, i.e., for the mean $\langle Z(t)\rangle$. Specifically, let
$\U(t,s)$ be the unique solution of
\beq
\partial_t\U(t,s)=\Q\Vt(t)\U(t,s)~~~\textrm{with}~~~ \U(s,s)=\mathsf{1}.
\label{eq:U}
\ee
Then,
\beq
\QZ(t)=\int_0^t ds\U(t,s)\Vt(s)\PZ(s)
\ee obeys (\ref{eq:QZ}) for the initial condition $\QZ(0)=0$ imposed by the initial condition (\ref{eq:ini-y}). Inserting this expression into (\ref{eq:PZ}), we get
\beq\partial_t \PZ(t)=\P\Vt(t)\int_0^tds \U(t,s)\Vt(s)\PZ(s).
\ee Finally, restoring $Y$ for $Z$ and $\V$ for $\Vt$, and using that $\Ab$ and $\P$ commute leads to 
\beq
\partial_t \P Y(t) = \Ab\P Y(t) 
 + \int_0^t ds \K(t,s)\P Y(s)  .
\ee
with the integral kernel
\beq
\K(t,s)\equiv  \P \V(t) e^{\Ab t}\U(t,s)e^{-\Ab s} \V(s) .
\label{eq:K}
\ee
As shown in Appendix \ref{appb}, the stationarity of the $\kappa(t)$-process implies that the integral kernel is time-translation invariant, i.e.,   $\K(t,s)=\K(t-s,0)$. 
Moreover, since the expression for $\K$ is bordered by two $\V$ matrices for which all entries but the 21-entry vanish, this matrix operator has as only non-vanishing element the 21-entry, with its time-translation invariant version  denoted as $K_{21}(t-s)$ in the following.

The equations of motion for the mean values $y_m(t)\equiv \langle  y(t) \rangle$ and $w_m(t)\equiv \langle \dot y(t)\rangle$, which are the two components of $\P Y(t)$, thus become
\beq
\partial_t\begin{pmatrix}
	y_m(t) \\
w_m(t)
\end{pmatrix} = \Ab \begin{pmatrix}y_m(t) \\w_m(t) \end{pmatrix}+ \begin{pmatrix}
	0 \\
	\int_0^t ds K_{21}(t-s) y_m(s)
\end{pmatrix}  .
\ee 
They are easily solved by  a Laplace transformation, $\hat f(z)\equiv \int_0^\infty dt e^{-zt} f(t)$, yielding
\beq
\begin{pmatrix}\hat y_m(z) \\ \hat w_m(z)  \end{pmatrix}= 
\frac{1}{D(z)}
\begin{pmatrix}z+1& 1\\ -\bar k +\hat K_{21}(z)& z \end{pmatrix}
\begin{pmatrix}y_m(0) \\ w_m(0)  \end{pmatrix}
\ee
with\beq D(z)\equiv {z(z+1)+\bar k - \hat K_{21}(z)} .
\ee The initial condition (\ref{eq:ini-y}) implies $y_m(0)=0$ and $w_m(0)=1$. 
With this Laplace transformation, the expression (\ref{eq:R3}) for $\mud$ becomes
\beq
\mud=\lim_{t\to \infty}y_m(t)=\lim_{z\to 0}z\hat y_m(z)= \lim_{z\to 0} z/D(z).
\ee

Since the differential mobility $\mud$ exists and is given by the long-time limit (\ref{eq:R3}), the latter implies that the Laplace transformation $\hat y_m(z)$ has to exist for all $z>0$. This fact implies that $D(z)\not= 0$ for all $z>0$. Moreover, for large $z$, $D(z)\approx z^2$ since $\lim_{z\to \infty}\hat K_{21}(z)=0$ due to $K_{21}(0)=\mathcal{O}(1)$ and $K_{21}(t)$ being continuous for small $t$. Hence $D(z)>0$ for $z>0$, since otherwise there was a $z^*>0$ with $D(z^*)=0$. Based on these properties, we can distinguish two cases.

 First, if $D(0)=\bar k - \hat K_{21}(0) > 0$, $\mud=0$, which includes the  case of a sufficiently strong non-periodic binding potential, where applying a force does not lead to a non-vanishing mean velocity. Second, and this is relevant for the periodic case in focus here, $D(z)$ may have a single root at $z=0$ with $D'(0)>0$, which implies
\beq
\bar k = \hat K_{21}(0)= \int_0^\infty dt  K_{21}(t)
\label{eq:id}
\ee and, finally, 
\beq
\mud=1/D'(0)=1/[1-\hat K_{21}'(0)]>0.
\label{eq:mud}\ee
Note that a root of higher than first order is excluded by the assumption that $\mud$ exists.

In conclusion, we have thus proven that the differential mobility cannot be negative for an underdamped one-dimensional Langevin dynamics in a periodic potential with a constant driving force. Moreover,
through this transformation to an oscillator, we have found a 
representation of the differential mobility through  an infinite series of higher-order dynamical curvature correlation functions. Whether this series expansion becomes useful for approximating or further analyzing the differential mobility remains to be seen.
\ack
I thank D. Klein for exploring a related approach numerically during his master thesis in my group. An interaction with Google Gemini (Free version) has inspired both the projection operator approach to (\ref{eq:Z})  and the proof of the time-translation invariance of the kernel in Appendix \ref{appb}.  

\appendix

\section{Overdamped case}
\label{appa}

In the overdamped case, a -- to the best of my knowledge -- novel, manifestly non-negative expression for the differential mobility, can be derived with a similar reasoning as follows.
In suitably scaled units, the overdamped Langevin equation
\beq\dot x = f-V'(x)  + \eta,
\ee
with $\langle \eta(t)\eta(t')\rangle = 2 \delta(t-t')$ implies the differential mobility
\beq\mud=\partial_f\langle \dot x\rangle^s = 1- \partial_f\langle V'(x)\rangle^s.
\label{eq:mudx}
\ee
The adjoint Fokker-Planck equation becomes
\beq
\partial_tp(x,t|x_0)= 
\LL^\dagger_{x_0} p(x,t|x_0)= \{[f-V'(x_0)]\partial_{x_0} + \partial^2_{x_0}\} p(x,t|x_0).
\label{eq:fpx-adj} 
\ee
A multiplication with $V'(x)-\overline {V'(x)}$, with $\overline {V'(x)}\equiv \langle V'(x)\rangle^s$, and an integration over $x$ yields
\beq
\partial_t\langle [V'(x)-\overline{V'(x)}|x_0\rangle = 
  \LL^\dagger_{x_0} \langle V'(x(t))-\overline{V'(x)}|x_0\rangle,
\ee 
where $\langle ...|x_0\rangle$ denotes an average over the noise at fixed initial value $x_0$.
Integrating over $t$ from 0 to $\infty$ and then operating with $\partial_f$ from the left yields 
\beq
\partial_f \overline{V'(x)}=[\partial_{x_0} +  \LL^\dagger_{x_0}\partial_f] 
\int_0^\infty dt  \langle V'(x(t))-\overline{V'(x)}|x_0\rangle.
\ee Finally, after multiplying this equation from the left with $p^s(x_0)$ and integrating over $x_0$, we get for the differential mobility (\ref{eq:mudx})  the expression
\beq \mud= 1-\langle \partial_{x_0} \int_0^\infty dt \langle V'(x(t))-\overline{V'(x)}|x_0 \rangle \rangle^s .
\label{eq:R1x}
\ee 
As in the main part, this derivative can be evaluated by considering two nearby trajectories with $x_\eps(t)=x(t)+\eps y(t)$ with $y(0)=1$ and the same noise history. The crucial derivative becomes
 \beq
 \partial_{x_0} \int_0^\infty dt  [V'(x(t))-\overline{V'(x)}] =\lim_{\eps\to 0}(1/\eps)\int_0^\infty dt V'(x(t)+\eps y(t))-V'(x(t))=\int_0^\infty dt V''(x(t)) y(t)
 \label{eq:partx}
 \ee
 where $y(t)$ solves 
 \beq\dot y(t)=-V''(x(t)) y(t)\ee with the initial condition $y(0)=1$.
If the solution $y(t)=\exp[-\int_0^t dt' V''x(t')]$ is inserted into (\ref{eq:partx}), the integrand becomes a total derivative that can be evaluated at the boundaries (with $T\gg1$ replacing $\infty$ for the moment) as
\beq
\partial_{x_0} \int_0^T dt \langle V'(x(t))-\overline{V'(x)}|x_0 \rangle =\big[-\exp[-\int_0^t dt' V''(x(t'))]\big]_0^T= 1-\exp[-\int_0^T dt V''(x(t))] .\ee
Inserting this into (\ref{eq:R1x}), averaging over the noise and the initial value $x_0$, and shifting $T\to \infty$ yields for the differential mobility
\beq \mud=\langle \exp[-\int_0^\infty dt V''(x(t))]\rangle,
\ee which is manifestly non-negative for each individual trajectory $x(t)$ even before averaging. 

\section{Time-translation invariance of the integral kernel}
\label{appb}
The propagator $\U(t,s)$ associated with the projected orthogonal dynamics can be expressed as a series expansion through integration of (\ref{eq:U}) and iterative  insertion as 
\begin{eqnarray}
\U(t,s)&=&\mathsf{1}+\int_s^t dt_1 \Q\Vt(t_1)\U(t_1,s)\\
&=&\mathsf{1}+\int_s^t dt_1 \Q\Vt(t_1)[\mathsf{1}+ \int_s^{t_1} dt_2\Q\Vt(t_2)\U(t_2,s)]\\
&=&1+\sum_{n=1}^\infty\U^{(n)}(t,s) 
\end{eqnarray}
with
\begin{eqnarray}
\U^{(n)}(t,s)&\equiv& \int_s^t dt_1 \int_s^{t_1} dt_2...\int_s^{t_{n-1}} dt_n \Q\Vt(t_1)...\Q\Vt(t_n)\\
&=&
\int_s^t dt_1 \int_s^{t_1} dt_2...\int_s^{t_{n-1}} dt_n \Q e^{-\Ab t_1}\V(t_1)e^{\Ab t_1}...\Q e^{-\Ab t_n}\V(t_n) e^{\Ab t_n}.
\label{eq:Un}
\end{eqnarray}
This series implies one for the integral kernel (\ref{eq:K})
\begin{eqnarray}
	\K(t,s)= \P \V(t) e^{\Ab t}\mathsf{1}e^{-\Ab s} \V(s) +
	\sum_{n=1}^\infty\K^{(n)}(t,s) 
\end{eqnarray}
with 
\beq
\K^{(n)}(t,s)=  \P \V(t) e^{\Ab t}\U^{(n)}(t,s)e^{-\Ab s} \V(s) .
\label{eq:Kn}
\ee
In the following, we will use that $\P$ and $\Q$ commute with the matrices $\Ab$ and $\V_0$. Inserting the propagator  (\ref{eq:Un}) into (\ref{eq:Kn}) then yields
\beq
\K^{(n)}(t,s)=  \P \V(t)
\int_s^t dt_1 \int_s^{t_1} dt_2...\int_s^{t_{n-1}} dt_n e^{\Ab(t-t_1)}\Q\V(t_1)e^{\Ab (t_1-t_2)}...\Q \V(t_n) e^{\Ab (t_n-s)} \V(s) .
\ee
We now insert (\ref{eq:V}) for $\V$ and pull all terms with $\Ab$ and $\V_0$ to the left, which yields
\begin{eqnarray}
\K^{(n)}(t,s)&=&  \V_0
\int_s^t dt_1 \int_s^{t_1} dt_2...\int_s^{t_{n-1}} dt_n e^{\Ab(t-t_1)}\V_0e^{\Ab (t_1-t_2)}...\V_0 e^{\Ab (t_n-s)} \V_0\nn \\&~&
\P\kappa(t)\Q\kappa(t_1)...\Q\kappa(t_n)\kappa(s)
\label{eq:K0}
\end{eqnarray}

In order to show time-translation invariance, we evaluate 
\begin{eqnarray}
	\K^{(n)}(t+\tau,s+\tau)&=&  \V_0
	\int_{s+\tau}^{t+\tau} dt_1 \int_{s+\tau}^{t_1} dt_2...\int_{s+\tau}^{t_{n-1}} dt_n e^{\Ab(t+\tau-t_1)}\V_0e^{\Ab (t_1-t_2)}...\V_0 e^{\Ab (t_n-s-\tau)} \V_0\nn\\&~&
	\P\kappa(t+\tau)\Q\kappa(t_1)...\Q\kappa(t_n)\kappa(s+\tau) .
\end{eqnarray}
 With the shifted integration variables $u_i\equiv t_i-\tau$, this expression becomes
 \begin{eqnarray}
 	\K^{(n)}(t+\tau,s+\tau)&=&  \V_0
 	\int_{s}^{t} du_1 \int_{s}^{u_1} du_2...\int_{s}^{u_{n-1}} du_n e^{\Ab(t-u_1)}\V_0e^{\Ab (u_1-u_2)}...\V_0 e^{\Ab (u_n-s)} \V_0\nn\\&~&
 	\P\kappa(t+\tau)\Q\kappa(u_1+\tau)...\Q\kappa(u_n+\tau)\kappa(s+\tau) .
\label{eq:K1}
 \end{eqnarray}
Stationarity of all multi-time correlations functions involving $\kappa(t)$ implies
 \beq
 \P\kappa(t+\tau)\Q\kappa(u_1+\tau)...\Q\kappa(u_n+\tau)\kappa(s+\tau)
 =
 \P\kappa(t)\Q\kappa(u_1)...\Q\kappa(u_n)\kappa(s).
 \ee Inserting this expression into (\ref{eq:K1}) and then setting $u_i=t_i$ shows that
 (\ref{eq:K1}) is equal to  (\ref{eq:K0}) for all $n$. Thus $\K(t+\tau,s+\tau)=\K(t,s)$, which proves time-translation invariance of the integral kernel.
%
%
%
\newpage



\providecommand{\newblock}{}

\end{document}